\documentclass[twocolumn, prl, amsmath,amssymb]{revtex4-1}
\usepackage[pdftex]{graphicx}
\usepackage{dcolumn}
\usepackage{bm}
\usepackage[T1]{fontenc}
\usepackage[utf8]{inputenc}
\usepackage{natbib}
\usepackage{hyperref}
\usepackage{xcolor}
\usepackage{url}
\usepackage{ulem}

{\rule{20pt}{1ex}\endlist}
\let\oldmarginpar\marginpar
\renewcommand\marginpar[1]{\-\oldmarginpar[\raggedleft\footnotesize #1]%
{\raggedright\footnotesize #1}}

\begin{document}

\newcommand{\al}{\alpha}
\newcommand{\ga}{\gamma}
\newcommand{\De}{\Delta}
\newcommand{\mcal}[1]{\mathcal{#1}}
\newcommand{\myref}[1]{(\ref{#1})}
\newcommand{\dd}{{\rm d}}
\newcommand{\sur}[2]{{\displaystyle\mathop{#1}_{#2}}}
\newcommand{\demi}{\frac{1}{2}}
\newcommand{\de}{\delta}

\title{Elastocapillary deformation of thin elastic ribbons in 2D foam columns}
\author{Manon Jouanlanne$^{1}$}
\author{Antoine Egelé$^{1}$}
\author{Damien Favier$^{1}$}
\author{Wiebke Drenckhan$^{1}$}
\author{Jean Farago$^{1}$}
\author{Aur\'elie Hourlier-Fargette$^{1}$}
\affiliation{
$^1$ Université de Strasbourg, CNRS, Institut Charles Sadron UPR22, F-67000 Strasbourg, France\\}

\date{\today}

\begin{abstract}

The ability of liquid interfaces to shape slender elastic structures provides powerful strategies to control the architecture of mechanical self assemblies. However, elastocapillarity-driven intelligent design remains unexplored in more complex architected liquids - such as foams. Here we propose a model system which combines an assembly of bubbles and a slender elastic structure. Arrangement of soap bubbles in confined environments form well-defined periodic structures, dictated by Plateau's laws. We consider a 2D foam column formed in a square section cylinder in which we introduce an elastomer ribbon, leading to architected structures whose geometry is guided by a competition between elasticity and capillarity. In this system, we quantify both experimentally and theoretically the equilibrium shapes, using X-ray micro-tomography and energy minimisation techniques. Beyond the understanding of the amplitude of the wavy elastic ribbon deformation, we provide a detailed analysis of the profile of the ribbon, and show that such setup can be used to grant a shape to a UV-curable composite slender structure, as a foam-forming technique suitable to miniaturisation. In more general terms, this work provides a stepping stone towards an improved understanding of the interactions between liquid foams and slender structures.

\end{abstract}

\maketitle

\section{Introduction}

\begin{figure}[h!]
\centering
\includegraphics{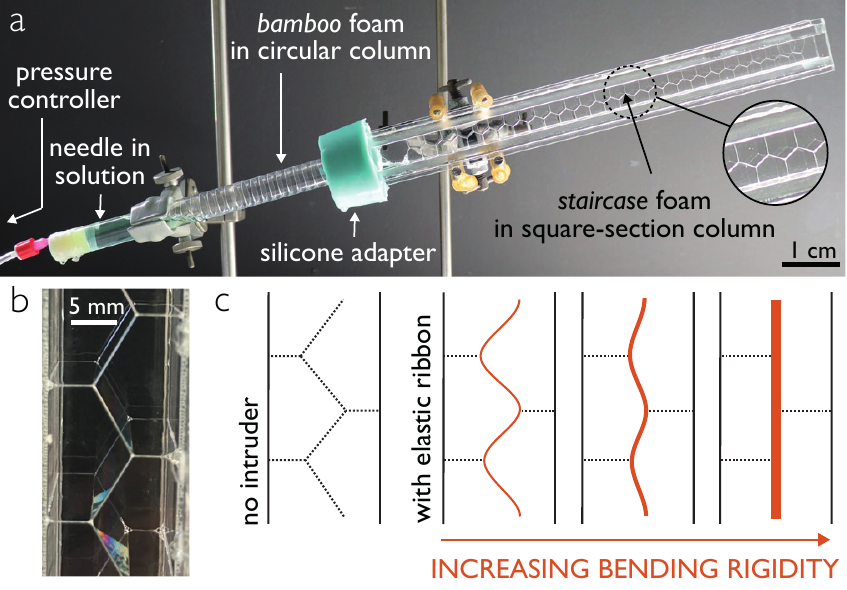}
\caption{\textbf{Experimental setup} (a) Bubbles generated by blowing air through a needle at constant pressure into a detergent solution form a bamboo foam in a circular column, transitioning to a staircase structure in a square cross-section column (b). Note the invariance by translation along the axis perpendicular to the column, providing a quasi-2D structure. (c) Overview of the study: the initial foam structure with no intruder (left) - dictated by Plateau's rules - is modified upon introduction of an elastic ribbon (right), with an equilibrium shape dependent on the bending rigidity of the ribbon.}
\label{fig:setup}
\end{figure}

Combination of soft materials and fluids offer a rich physics in which both elasticity and capillarity come into play \cite{Bico2018}, opening an area of opportunity for novel materials and fabrication strategies in the case of slender elastic structures \cite{Roman2010} and of bulk systems \cite{Style2020}. Geometry and size effects play a key role in the control of the shape of slender solids by capillary effects, for which bending dominates over stretching \cite{Holmes2019}. Nature provides numerous examples such as the aggregation of wet hair \cite{Bico2004} or the spooling of spider web in liquid droplets \cite{Elettro2016}, serving progressively as a source of inspiration for the design of innovative materials \cite{Grandgeorge2018}. Turning catastrophic events such as capillarity-induced collapse \cite{Lau2003} into robust microfabrication techniques \cite{Syms2003,De-Volder2013} has promissing impacts in electronics and energy harvesting, soft robotics, or even drug encapsulation and delivery \cite{Hines2016,Kwok2020}. We consider here the case of a slender structure introduced in an architected liquid, formed by a quasi-2D column of bubbles (Fig. \ref{fig:setup}). Mechanical self-assembly of soap bubbles provides foam structures that - in the  limit of low liquid content - obey specific geometrical and topological rules dictated by Plateau's laws \cite{Plateau1873}. Further confinement of bubbles in columns of square-cross section results in well-defined ordered structures, provided that the enclosure dimensions are comparable to the bubble diameters \cite{Hutzler2009}. Such systems form periodically ordered liquid film architectures into which an elastic ribbon can be introduced. Although the question of complex deformation of an intruder in an architected medium has been widely studied in the case of granular media \cite{Mojdehi2016,Seguin2018} with applications in the context of root growth \cite{kolb2017}, a limited number of studies on model systems exist in the case of liquid foams, such as the analysis of simple arrangements of soap films interacting with rigid solids \cite{Whyte2017}, and simulations of flexible fibres in foam under shear \cite{Langlois2017}.  

We consider a model system composed of an ordered assembly of monodisperse soap bubbles and an elastomer ribbon (Fig. \ref{fig:setup}). Among the different possible arrangements governed by the confinement ratio of the bubbles inside the column, we select the so-called \textit{staircase} structure \cite{Reinelt2001,Hutzler2009}  (Fig. \ref{fig:setup}b) that offers an invariance by translation along the axis perpendicular to the column, providing a quasi-2D model system.

The staircase structure presents a central soap film composed of straight sections connected at 120$^{\circ}$ angles (Fig. \ref{fig:setup}c left) into which we insert an elastic intruder  (Fig. \ref{fig:setup}c right). Depending on the bending rigidity of the elastomer ribbon, the resulting shape of the structure evolves from a case close to the geometry with no intruder to a system where two \textit{bamboo foam} columns (equally spaced parallel soap films \cite{Hutzler2009}) are separated by a flat plane, as illustrated with the orange arrow on Fig. \ref{fig:setup}c. In the following, we provide an experimental and theoretical framework to quantify such equilibrium shapes, both showing excellent agreement. \\

\section{Experimental methods}
\subsection{Materials}
All experiments are carried out with an aqueous solution of 4.5 vol.\% commercial detergent (Fairy Liquid), 1.5 vol.\% Glycerol (Sigma - Aldrich) and 10 g/L industrial PEO lubricant Jlube (Jorgensen Labs) \cite{Frazier2020}, resulting in bubble structures which are stable over tens of minutes under the experimental conditions of this study (Fig. S1 in the supplementary material). Water, PEO and glycerol are mixed with a magnetic stirrer for several hours, and the detergent is added after 24 h. Solutions are used only after an additional 24 h period of rest. The surface tension of this solution, measured using a Kibron tensiometer, is found to be $\gamma = 26.5 \pm 0.5$ mN/m.
 The elastic ribbons are made of polydimethylsiloxane (PDMS, Dow Corning Sylgard 184 Elastomer base mixed with its cross-linker in proportion 10:1). They are cut to a width $w = 14.5 \pm 0.1$ mm from 100 x 100 mm films of variable thickness $t$ produced with a Laurell WS-650MZ spin coater at speeds ranging from 200 to 2000 rpm, subsequently cured in an oven at 60$^{\circ}$C for two hours and stored for at least 48h at room temperature before performing the experiments. The thicknesses $t$ of the ribbons are measured with a Bruker optical profilometer. The Young’s modulus of PDMS is $E = 1.7 \pm 0.2$ MPa (\cite{Johnston2014}, confirmed by our own DMA testing on a 1500 rpm sample) and the Poisson's ratio $\nu$ is taken as equal to 0.45. Both sides of the ribbon are hydrophilised via plasma cleaner treatment (Harrick Plasma) at high intensity for 1 min (Fig. S2 in the supplementary material).
 
\subsection{Experimental setup}

Bubble columns are generated in a circular perspex tube (ID 16 mm) sealed on one side with a cork that includes a nozzle (ID 0.61 mm) near the tube wall (Fig. \ref{fig:setup}a) \cite{Elias2007}. Using a silicone adapter, we connect to the first tube a PMMA column with a 15 mm-square cross-section into which we place a stretched wire at the end of which an elastic ribbon is suspended.
We fill the circular tube with the aqueous detergent solution and tilt it by approximately 10 degrees so that the nozzle opening is just below the air/liquid interface \cite{Elias2007}. Air is blown into the solution at a constant pressure (10 mbar) via an Elveflow pressure controller (OB1 MkII), resulting in the formation of a \textit{bamboo structure}. These equally spaced soap films slowly slide upwards through the silicone adapter and into the square section column where the soap films rearrange into a \textit{staircase structure}. 
We then disconnect the square section column from the bubbling setup and introduce the ribbon into the central soap films by carefully pulling on the wire (Fig. S3 in the supplementary material). We close the column on the top side with a silicone plug, leaving the other side open to facilitate drainage of liquid out of the foam structure for 30 min. The resulting system is stable for another 15 minutes (Fig. S1 in the supplementary material) during which shape and position of the ribbon remain fixed, which permits 3D scanning of the system with a X-ray microtomograph EasyTom 150/160 without motion artefacts. The tomographic images are made with resolutions ranging from 12 to 58.8 $\mu$m.

\begin{figure}[h!]
\centering
\includegraphics[width=86mm]{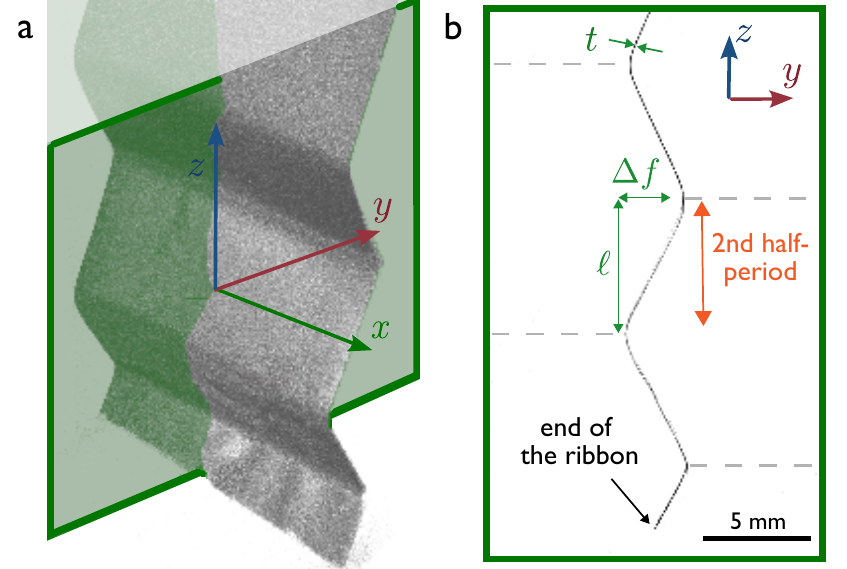}
\caption{\textbf{X-ray tomography measurements} (a) 3D-view of the elastomer ribbon in the soap bubble column after 30 minutes of drainage. Note that only the elastic ribbon is visible in the scan, as the soap film thicknesses are smaller than the voxel sixe. (b) Slice perpendicular to the $x$ axis (green color code in (a)), with definition of the measured amplitude $\Delta f$ and half-period $\ell$, highlighting the region of interest (second half-period from the end of the ribbon). Note that the thickness $t$, mentioned on the figure, is actually measured via an optical profilometer.}
\label{fig:tomo}
\end{figure}

Tomography provides slices across the $x$ axis corresponding to the width of the ribbon (Fig. \ref{fig:tomo}a). On each slice, we measure the amplitude of the ribbon $\Delta f$ and the height of the second half-period $\ell$ (Fig. \ref{fig:tomo}b) using a purpose-designed MATLAB code. To account for small deviations from a perfect 2D system, we calculated the average for $\Delta f$ and $\ell$ over 100 equally spaced slices along the $x$ axis.  We made all our measurements on the second half-period first to avoid edge effects, but also because higher half periods are prone to a flattening due to the weight of lower parts of the ribbon \cite{miller2014}, an effect disregarded in the theoretical treatment for sake of simplicity.

\section{Results and discussions}

\subsection{Experimental results}

\begin{figure*}
\centering
\includegraphics[width=17.2cm]{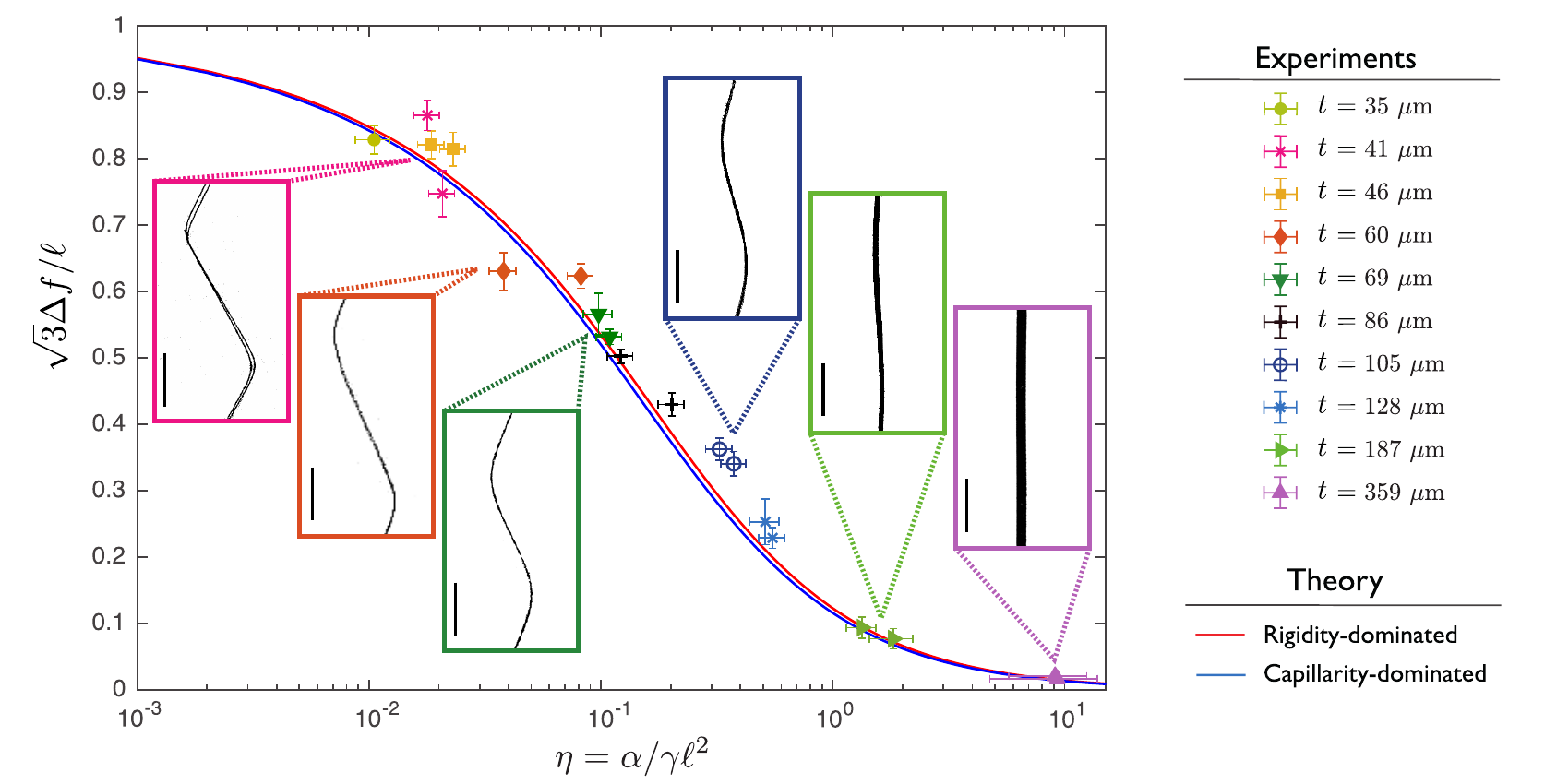}
\caption{\textbf{Deflexion measurements} Dimensionless deflection $\sqrt{3}\Delta f /\ell$ as a function of the parameter $\eta= \al /(\ga\ell^2)$ comparing bending rigidity of the ribbon and capillarity. The left side of the graph corresponds to flexible ribbons, the right side to rigid ribbons. Theoretical prediction for rigidity-dominated case (quadratic order around the \textit{straight ribbon} configuration) and for capillarity-dominated case (quadratic order around the \textit{bubble-only} configuration) are shown with straight lines, respectively in red and blue. Experiments correspond to the second half-period of PDMS ribbon with 10 different thicknesses, resulting in different bending rigidities. Error bars correspond to both systematic uncertainties on $\alpha$ and $\gamma$ and statistical variation of the parameters $\Delta f$ and $\ell$ over 100 slices taken along the $x$ axis, covering the full ribbon width. Insets show X-ray tomographies of corresponding systems (scale bars are 2 mm).}
\label{fig:mastercurve}
\end{figure*}

We focus on the dependence of the equilibrium shapes of ribbon-foam couples as a function of the competition between elasticity and capillarity in such systems, by introducing into the foam PDMS ribbons of 10 different thicknesses, ranging from 35 $\mu$m to 359 $\mu$m. This represents a variation of the bending rigidity of the ribbons over three orders of magnitude. Among the other parameters in the system, the width and Young's modulus of the ribbon, the dimensions of the square section column and the surface tension are kept constant, while the  half-period $\ell$ is measured on each bubble column and comprised between $4.2 \pm 0.1$ and $6.2 \pm 0.2$ mm. Qualitatively, the stiffer the ribbon the smaller the amplitude $\Delta f$ (Fig. \ref{fig:setup}c). 
We present in Fig. \ref{fig:mastercurve} the dimensionless deformation of PDMS ribbons of thicknesses $t=$ 35, 41, 46, 60, 69, 86, 105, 128, 187 and 359 $\mu$m as a function of the dimensionless bending rigidity $\eta$, together with profiles of the corresponding ribbons captured via X-ray tomography. For the deformation of the ribbon, we consider the parameter $\sqrt{3} \Delta f / \ell$, which is equal to 1 in the limit case of soft ribbons where the geometry is dictated by Plateau's laws. On the abscissa, the dimensionless bending rigidity $\eta$ is defined as 

\begin{align}
    \eta&= \frac{\al}{\ga\ell^2},
\end{align}

\noindent where $\alpha$ is the bending rigidity per unit width of the ribbon and $\gamma$ the surface tension. The parameter $\eta$ compares the elastocapillary length $\sqrt{\al/\ga}$ to the geometrical length $\ell$ of the problem. For length scales $\ll \sqrt{\al/\ga}$ (resp. $\gg$), the physics is mainly dictated by the elasticity (resp. the capillarity). \\

For large values of $\eta$, the deformation of the ribbon is small, and increases when decreasing $\eta$. At the limit $\eta \ll 1$, the shape is close to the initial pattern of bubbles, with angles prescribed by Plateau's laws. To go beyond the description of the deformation in terms of amplitudes, we also extract the full profiles, presented in a dimensionless manner in Fig. \ref{fig:shapes}, for PDMS ribbons of various thicknesses spanning the range shown in Fig. \ref{fig:mastercurve}. In order to rationalise these results, the following section presents the theoretical modeling of both amplitudes and shapes of the profiles, to which our experimental results will be compared.  \\

\subsection{Theoretical interpretation}
 
The equilibrium state of the ribbon in the staircase bubble assembly can be obtained by minimising the free energy of the system. The excess free energy (per unit length in the direction transverse to the Fig. \ref{fig:setup}.c) of the system with respect to the situation where only the staircase bubbles are present is given by
\begin{align}
    \mcal{F}&=\mcal{E}_{\rm ribbon}+\De\mcal{E}_{\ga} +\De F_{\rm gas},\label{F}
\end{align}
where $\mcal{E}_{\rm ribbon}$ is the elastic energy per unit width of the ribbon, $\De\mcal{E}_{\ga}$ is the excess interfacial energy per unit width of the liquid films, with respect to the situation without ribbon. Notice that we assume that the ribbon is everywhere in the {\em interior} of the liquid films so that the area of the ribbon itself contributes to the liquid interfacial energy. The final term $\De F_{\rm gas}$ accounts for the free energy cost of the possible contraction/dilation of the gas within the bubbles due to the presence of the ribbon, and is neglected in the following, as justified in the Appendix.

The actual expression for the elastic term is
\begin{align}
    \mcal{E}_{\rm ribbon}&=\frac{\al}{2}\int C^2\dd s\label{pouet},
\end{align}
where $\al=Et^3/12 (1-\nu^2)$, and $s$ is the curvilinear abscissa along the ribbon, related to the ribbon profile $f(z)$ by $\dd s=\dd z\sqrt{1+f'(z)^2}$. The curvature $C$ is given in terms of $f(z)$ by
\begin{align}
    C&=\frac{f''(z)}{\sqrt{1+(f'(z))^{3/2}}}\label{C}.
\end{align}
For a long and uniform ribbon, which zigzags periodically along $N\gg 1$ identical bubbles (we count the bubbles of length $2\ell$ only on one side of the ribbon (Fig. \ref{fig:tomo}b)), we can write, up to negligible boundary corrections, $\mcal{E}_{\rm ribbon}\simeq \frac{N\al}{2}\int_{\rm 1p} C^2 \dd s$, where $\int_{\rm 1p}(\ldots)$ assumes an integration over one period of the ribbon only.

Making similar assumptions, the second term of the energy in Eq. \myref{F} is written 
\begin{align}
    \De\mcal{E}_{\ga}&\simeq 2\ga N\left[\int_{\rm 1p}\dd s-\De f-\sqrt{3}\ell \right]\label{pouet2},
\end{align}
where $\De f=\max f(z)-\min f(z)$ is the transverse amplitude of the ribbon (Fig. \ref{fig:tomo}). Note that (i) the factor 2 corresponds to the two liquid-air interfaces of the liquid films, (ii) the last term of Eq. \myref{pouet2} comes from the fact that the zero energy reference state is chosen to be the column of bubbles without ribbon 
for which the interfacial energy per unit width over one period is $2 \gamma (\sqrt{3}\ell+w)$ (assuming the square column width and the ribbon width to be equal), and (iii) $w$ does not appear in Eq. \myref{pouet2} due to the translational invariance of the bubble pattern along the $y$-direction. Moreover, as explained in the Appendix, the half length of the bubble $\ell$ can be considered constant during the minimization process, because the compression/dilation of the gas due to the ribbon is negligible.
\bigskip

We consider now the minimisation process of the free energy : The equilibrium profile $f(z)$ is the one which minimises $\mcal{F}$ (Eq. \myref{F}). This minimisation is complex for two reasons. Firstly, since the functional $\mcal{F}$ is not quadratic in $f(z)$, we will consider two different quadratic approximations according to the physical properties of the ribbon (flexible and rigid limit cases), to obtain a tractable theory. Secondly, for a long homogeneous ribbon, we anticipate that the equilibrium shape is periodic with $N$ identical oscillations embedded in a deformed network of bubbles. But, in contrast to the total length $L$ of the ribbon, the number $N$ is not a constant of the minimisation procedure, since the deformation of the ribbon reduces the number $N$ of oscillations the ribbon can develop. To tackle this specific difficulty, two equivalent routes can be followed. Lagrange multipliers could be used to account for the total length of the ribbon, the number $N$ being temporarily treated as a constant. Alternatively --- and it is the route we follow here ---, the explicit relation $L\simeq N\int_{\rm 1p}\dd s$ allows to both account for the constancy of the total length of the ribbon and the variation of $N$. As a result, the unconstrained functional to minimize becomes
\begin{align}
    \mcal{F}_1&\equiv \frac{\mcal{F}}{\ga L}-2=\frac{\frac{\al}{2\ga}\int_{\rm 1p} C^2\dd s-2[\De f+\sqrt{3}\ell]}{\int_{\rm 1p}\dd s}.
    \label{F1_def}
\end{align}
Taking into account the internal mirror symmetry of the expected optimal profile (Fig. \ref{fig:setup}), Eq. \myref{F1_def} can be rewritten
\begin{align}
    \mcal{F}_1&=\frac{\al}{2\ga}\frac{\int_0^\ell \sqrt{1+(f')^2}C^2\dd z}{\int_0^\ell\sqrt{1+(f')^2}\dd z}-\frac{\De f+\sqrt{3}\ell}{\int_0^\ell\sqrt{1+(f')^2}\dd z},\label{F1}
\end{align}
with $\De f=f(\ell)-f(0)$ if $z=0$ is taken at the position of a transverse liquid film, such that $f(0)=\min f$ (this value being arbitrary, we choose $f(0)=0$ in the following). For pure bubbles, i.e. in the limit $\al\rightarrow 0$, the optimal $f(z)$ given by Eq. \myref{F1} is $f(z)=z/\sqrt{3}$, in accordance with Plateau's laws \cite{Plateau1873}, which prescribe 120$^\circ$ angles between connecting films.

\medskip

The physics of the problem is governed solely by the dimensionless parameter
\begin{align}
    \eta&\equiv \frac{\al}{\ga\ell^2}.
\end{align}
For high values of $\eta$, i.e. for stiff ribbons, one expects that $f(z)/\ell \ll 1$ and a second order expansion in $f$  for $\mcal{F}_1$ is physically relevant. Disregarding irrelevant constants, the quadratic approximation for  $\mcal{F}_{1}$ for large $\eta$ reads
\begin{align}
    \mcal{F}_{1}&\sur{\simeq}{\eta\gg1}\frac{\eta\ell}2\int_0^\ell(f'')^2\dd z+\frac{\sqrt{3}}{2\ell}\int_0^\ell(f')^2\dd z-\frac{\De f}{\ell}. \label{F1quad}
\end{align}
We write $f(z)=(\De f)g(z/\ell)$ so that $g(0)=0$ and $g(1)=1$ and consider first the optimisation of the shape $g$ before considering the optimisation of the amplitude $\De f$. The optimal $g$ is found using standard techniques of Lagrangian mechanics \cite{Goldstein,Lanczos} (notice here that the four boundary conditions of the problem are $g(0)=0$, $g(1)=1$, $g'(0)=g'(1)=0$) 
\begin{align}
    g_{\rm opt}(z/\ell)&=\frac{\tanh(\kappa)\sinh^2(\kappa z/\ell)-[\demi\sinh(2\kappa z/\ell)-\kappa z/\ell]}{\kappa-\tanh(\kappa)},\label{onze} \\
   \text{with } \kappa&\equiv\frac{3^{1/4}}{2\sqrt{\eta}}\simeq \frac{0.658}{\sqrt{\eta}}. \label{kappa_large}
\end{align}

Once $g_{\rm opt}$ is known, $\Delta f$ is computed as the value minimizing Eq. \myref{F1quad}, which is simply a second order polynomial in $\De f/\ell$. One finds
\begin{align}
    \frac{\sqrt{3}\De f}{\ell}&\sur{\simeq}{\eta\gg 1}1-\frac{\tanh\kappa}{\kappa}.\label{Deltaf_large}
\end{align}
In the $\eta \gg 1$ regime, for which this formula is in principle only relevant, one can write one step further $\sqrt{3}\De f/\ell\simeq \kappa^2/3=\eta^{-1}/[4\sqrt{3}]$, showing that $\De f$ goes to zero as $\propto \al^{-1}$ for large $\al$. It is interesting to note that in the opposite range $\eta\rightarrow 0$, Eq. \myref{Deltaf_large}, though not supposed to work here, gives however the correct limit $=1$.

\bigskip

Actually the limit of small $\al$ (or small $\eta$) is incorrectly described by the previous theory, because  one expects here the central zigzag of the bubble pattern to be hardly perturbed by the ribbon, therefore assuming $\De f/\ell\ll 1$ is simply incorrect. The correct method is to write $f(z)=z/\sqrt{3}+\xi(z)$ and assume that $\xi/\ell\ll 1$. As $z/\sqrt{3}$ is the equation associated to half of a period of the central line of the bubble network (in absence of the ribbon), one expects that weak values of $\al$ will induce only minor departures from this pattern. The quadratic approximation of Eq. \myref{F1} in the field $\xi(z)$ can be re-expressed in the field $f(z)$ and reads (up to a constant) 
\begin{multline}
    \mcal{F}_{1}\sur{\simeq}{\eta\ll 1}\frac{\eta\ell}{2(1+3^{-3/4})}\int_0^\ell (f'')^2\dd z+\frac{9}{16\ell}\int_0^\ell(f')^2\dd z\\-\frac{3\sqrt{3}}{8}\frac{\De f}{\ell}.
\end{multline}

In this case, the optimal solution is given by formulas very similar to the preceding case. As before, one writes a priori $f(z)=(\De f)g(z/\ell)$ and finds respectively for $g$ and $\De f/\ell$ the results given in Eq. \myref{onze} and Eq. \myref{Deltaf_large} with $\kappa$ replaced by
\begin{align}
    \kappa_<&\equiv \frac{3(1+3^{-3/4})^{1/2}}{4\sqrt{2\eta}}\simeq\frac{0.636}{\sqrt{\eta}}. \label{kappa_small}
\end{align}
The two regimes are described by the same formulas, differing only by a modest change in the constant appearing in $\kappa$, a similarity which is somewhat surprising. Notice however that the profile shape given in Eq. \myref{onze} has a considerably different limiting expression for small values of $\eta$. For $\eta\gg1$, one can show that $g_{\rm opt}(z/\ell)\simeq  3(z/\ell)^2-2(z/\ell)^3$. For $\eta\ll1$ however, one has $g_{\rm opt}(z/\ell)\sim z/\ell$, a result expected because the bubble profile must be recovered at $\eta=0$. Notice that the latter limiting form is not compatible with the boundary conditions $g'=0$,  a discrepancy due the fact that a regularisation of the ribbon profile occurs near $z=0$ and $z=\ell$ over a length $\sim \ell/\kappa_<\sim \sqrt{\al/\ga}$, i.e. the elastocapillary length. As mentioned above, the elasticity dominates on length scales shorter than $\sqrt{\al/\ga}$, as exemplified here.

\begin{figure}[!h]
\centering
\includegraphics[width=86mm]{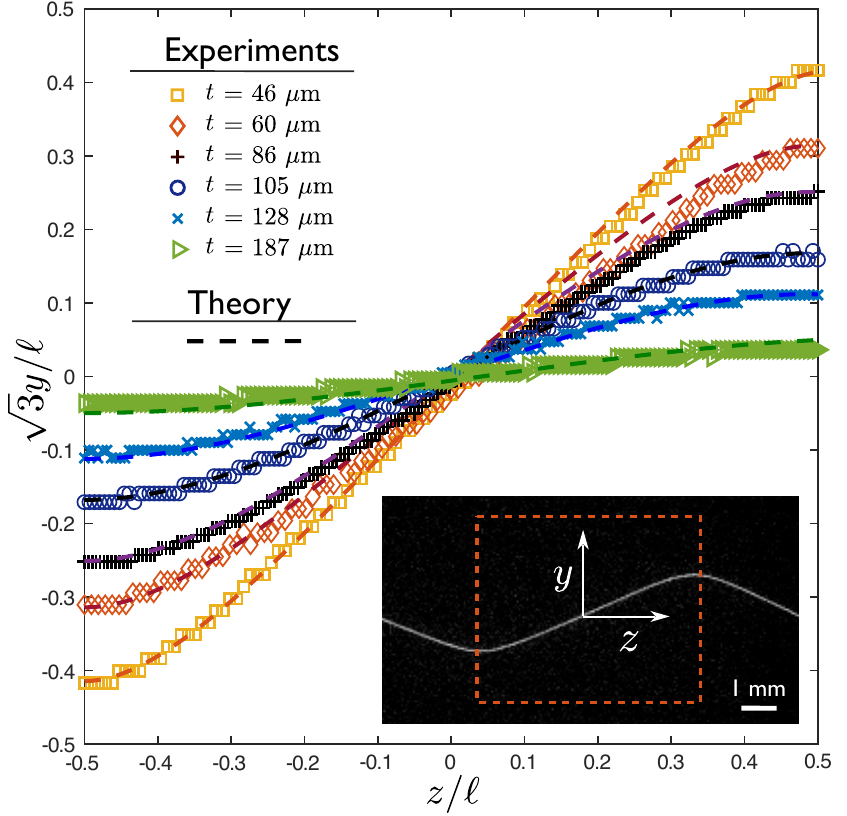}
\caption{\textbf{Equilibrium shapes} Dimensionless profile of ribbons of different thicknesses $t$ (and thus different bending rigidities), from flexible (yellow squares) to rigid (green triangles) structures. Theoretical shapes are calculated from the values of $\Delta f$, $\ell$ measured on the foam-ribbon system, and of $\eta$ measured on the ribbon, with no further fitting parameter, using the appropriate model (small deformation around straight line for rigid ribbons, small deformations around the bubble-only case for flexible ribbons). Inset: region of interest considered for those curves, and definition of $y$ and $z$ axis.}
\label{fig:shapes}
\end{figure}

We plot in Fig. \ref{fig:mastercurve} the dimensionless amplitude of the ribbon as a function of $\eta$, corresponding to Eq. \myref{Deltaf_large}, with values of $\kappa$ given by Eq. \myref{kappa_large} in the $\eta \gg 1$ limit (rigidity dominated) and Eq. \myref{kappa_small} in the $\eta \ll 1$ limit (capillarity dominated). We observe an excellent agreement between experiments and theory over the whole range of deformations, showing that the key ingredients chosen and the approximations made in the modeling in the two limit cases are relevant.

The full profile of the ribbons is described by Eq. \myref{onze}: Fig. \ref{fig:shapes} compares this theoretical prediction to experiments, using the values of $\Delta f$ and $\ell$ measured on the profile to establish Fig. \ref{fig:mastercurve}, and the value of $\eta$ measured from profilometry and Young's modulus, with no further fitting parameters. Again, our modeling captures well the experimental results, over the whole range of ribbon thicknesses.

\subsection{Application to bubble-based forming}

\begin{figure}[!h]
\centering
\includegraphics{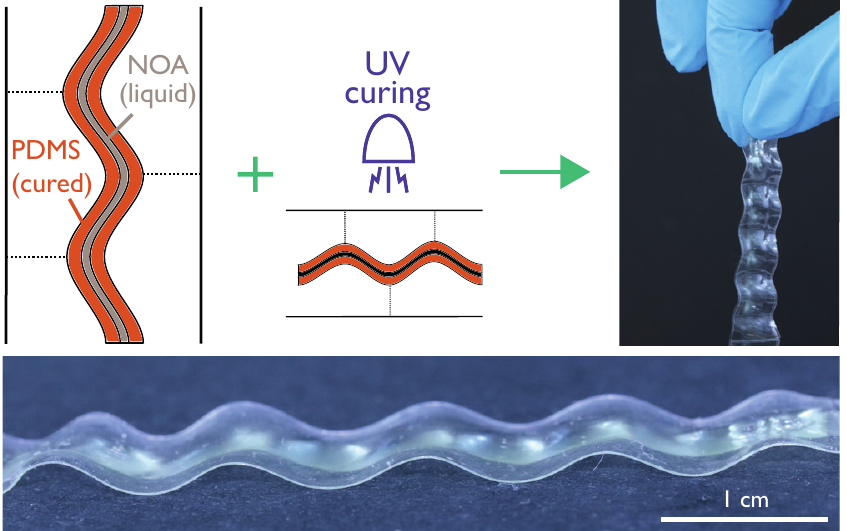}
\caption{\textbf{Application of foam-driven ribbon bending to UV-curable composite systems}. Bonding of two cured PDMS layers provides an outer shell in which a liquid UV-curable adhesive (NOA) is poured. This composite structure undergoes shaping through insertion inside the bubble column and subsequent UV curing. The resulting solid self-standing structure presents corrugations dictated by the competition between elasticity and capillarity in the uncured system.}
\label{fig:application}
\end{figure}

Beyond those results, we demonstrate that such setup can be used as a method to imprint shapes to materials by solidifying a UV-curable thin elastic ribbon in the bubble assembly. To do this, we prepare two crosslinked PDMS ribbons of $70 \pm 5\ \mu$m thickness and $14.5 \pm 0.1$ mm width, treated with plasma cleaner for 2 minutes at high intensity in order to activate their surfaces. A mask protects a rectangular part (8 x 100 mm) from the plasma, located at the middle of the ribbons. We put the two activated surfaces in contact, apply a light pressure by hand for 2 minutes and place it in the oven for 15 minutes at 60$^{\circ}$C. Plasma activation of the surfaces results in strong bonding of the two PDMS surfaces \cite{Eddings2008}. Only the edges of the ribbon (not protected by the mask) are glued together, so we obtain a PDMS shell which we fill with NOA 85 (Norland Optical Adhesive), a liquid photopolymer that cures when exposed to UV light. Following the same protocol as for PDMS ribbons, described in the experimental setup section and in Fig. S3 in the supplementary material, we slide the NOA-filled PDMS shell inside the staircase structure and place it in a Creality UW-01 curing machine during 10 minutes at low rotation speed. The result is a cured ribbon made of two PDMS layers with a NOA 85 layer in the middle, and shaped according to the characteristic geometries of the bubble-ribbon structure (Fig. \ref{fig:application}). In the UV-cured state, as NOA 85 Young's modulus is almost two orders of magnitude higher than PDMS, the thin layer of NOA is sufficiently stiff to maintain the wavy shape when the composite ribbon is extracted from the foam. With this simple proof-of-concept experiment, we demonstrate that a liquid foam can be used to shape slender objects, with a resulting corrugation imprinted by the foam and the competition between elasticity and capillarity in such systems.

\section{Conclusion}

In summary, we have highlighted that the equilibrium geometry of bubble columns can be tuned upon introduction of elastic ribbons, resulting in shapes that differ from classical foam structures dictated by Plateau's laws. We have shown that the resulting geometry of the ribbon-foam system is controlled by the competition between the bending energy of the ribbon and the interfacial energies of the bubble surfaces, as illustrated by the excellent agreement between our experiments and theory for both the amplitude and the full profile of the structures. Liquid foams are excellent model systems for complex cellular structures, and foams including elastic membranes as the one we describe here could have potential relevance for tissue growth \cite{Landsberg2009}. Interactions between foams and flexible slender structures also occur in widely used fabrication techniques such as foam-forming of cellulose for the paper industry \cite{Burke2021, Hjelt2021}. Finally, using a UV-curable system, we have proposed a way to mold materials with characteristic shapes and curves resulting from the elasto-capillary competition.

\section{Appendix: effect of the gas compressibility}

In this Appendix, we turn our attention to the last term of Eq. \myref{F}, which corresponds to a possible density change of the gas phase caused by the presence of the ribbon. Notice that this density change could occur only via a change of the parameter $\ell$ corresponding to half of a bubble longitudinal size (Fig. \ref{fig:tomo}) : Along the lateral dimension $y$ any imbalance of density along two oppositely placed bubbles would cause a pressure imbalance and thus a translation (at zero elasto-capillary energetic cost) of the bubble pattern to restore the pressure equality on each side of the ribbon. Neglecting again the boundary effects near the ends of the ribbon, one can write $\De F_{\rm gas}\simeq NF(\text{bubble+ribbon})+(N_{\rm tot}-N)F(\text{bubble outside ribbon})-N_{\rm tot}F(\text{pure bubble})$. If one assumes that the ribbon induces a modification of the half period $\ell+\de\ell$ with $\de\ell\ll \ell$, a lowest order expansion yields $\De F_{\rm gas}\simeq \frac{N(\de\ell)^2w}{\ell\chi_T}$. Despite $\De F_{\rm gas}$ is the free energy of the gas per width in the transverse $x$ direction, $w$ is however present in the formula due to the extension of the gas in the $y$ direction. If the column width was infinitely large, this relation shows that changes in $\ell$ would be simply forbidden because prohibitively costly. For finite $w$ (we have $w= 15$ mm in our experiment), we have to quantify the typical $\de\ell$ induced by the ribbon. The typical value of the elastic energy per unit width is dimensionally given by const$\times N\al /\ell$ because the only lengthscale of the problem is $\ell$, therefore a shift $\de\ell$ of the value of $\ell$ induces an elastic energy contribution $\sim-\al\de\ell/\ell^2$, and the value of $\de\ell$ is found when the thermodynamic forces induced by $\De F_{\rm gas}$ and this elastic term are equivalent. We get 
\begin{align}
    \frac{\de\ell}{\ell}\sim \frac{\al\chi_T}{w\ell^2}\sim \frac{E}{10 P_{\rm gas}}\left(\frac{t}{\ell}\right)^3,
\end{align}
where in the last relation we have considered that $w\sim \ell$. With the typical values used in our experiments, we have $E/P_{\rm gas}\simeq 10$ and $t/\ell\simeq 10^{-2}$ so that $\de\ell/\ell\simeq 10^{-6}$, i.e. the compression of the gas can be safely neglected and $\ell$ be considered as a constant. Pay attention however to the fact that $\ell$ varies from one experiment to another due to unavoidable variations during the bubble generation process.

\section{Acknowledgements}
We thank Thierry Charitat, François Schosseler, Léandro Jacomine, Sébastien Andrieux, Christian Gauthier and Sébastien Neukirch for fruitful discussions, ICS PECMAT team and especially Fouzia Boulmedais for the access to a spin-coater tool and to a plasma cleaner, Sébastien Kauffmann for suggestions of detergent solution formulation, Camille Kauffer for her work on preliminary experiments, and Christophe Lambour for the fabrication of the square-section column. This work of the Interdisciplinary Institute HiFunMat, as part of the ITI 2021-2028 program of the University of Strasbourg, CNRS and Inserm, was supported by IdEx Unistra (ANR-10-IDEX-0002) and SFRI (STRAT’US project, ANR-20-SFRI-0012) under the framework of the French Investments for the Future Program. We also acknowledge funding from the IdEx Unistra framework (Chair W. Drenckhan) and from the European Research Council (ERC- METAFOAM 819511).\\

\bibliographystyle{apsrev4-1}
\bibliography{references}

\begin{thebibliography}{30}%
\makeatletter
\providecommand \@ifxundefined [1]{%
 \@ifx{#1\undefined}
}%
\providecommand \@ifnum [1]{%
 \ifnum #1\expandafter \@firstoftwo
 \else \expandafter \@secondoftwo
 \fi
}%
\providecommand \@ifx [1]{%
 \ifx #1\expandafter \@firstoftwo
 \else \expandafter \@secondoftwo
 \fi
}%
\providecommand \natexlab [1]{#1}%
\providecommand \enquote  [1]{``#1''}%
\providecommand \bibnamefont  [1]{#1}%
\providecommand \bibfnamefont [1]{#1}%
\providecommand \citenamefont [1]{#1}%
\providecommand \href@noop [0]{\@secondoftwo}%
\providecommand \href [0]{\begingroup \@sanitize@url \@href}%
\providecommand \@href[1]{\@@startlink{#1}\@@href}%
\providecommand \@@href[1]{\endgroup#1\@@endlink}%
\providecommand \@sanitize@url [0]{\catcode `\\12\catcode `\$12\catcode
  `\&12\catcode `\#12\catcode `\^12\catcode `\_12\catcode `\%12\relax}%
\providecommand \@@startlink[1]{}%
\providecommand \@@endlink[0]{}%
\providecommand \url  [0]{\begingroup\@sanitize@url \@url }%
\providecommand \@url [1]{\endgroup\@href {#1}{\urlprefix }}%
\providecommand \urlprefix  [0]{URL }%
\providecommand \Eprint [0]{\href }%
\providecommand \doibase [0]{http://dx.doi.org/}%
\providecommand \selectlanguage [0]{\@gobble}%
\providecommand \bibinfo  [0]{\@secondoftwo}%
\providecommand \bibfield  [0]{\@secondoftwo}%
\providecommand \translation [1]{[#1]}%
\providecommand \BibitemOpen [0]{}%
\providecommand \bibitemStop [0]{}%
\providecommand \bibitemNoStop [0]{.\EOS\space}%
\providecommand \EOS [0]{\spacefactor3000\relax}%
\providecommand \BibitemShut  [1]{\csname bibitem#1\endcsname}%
\let\auto@bib@innerbib\@empty
\bibitem [{\citenamefont {Bico}\ \emph {et~al.}(2018)\citenamefont {Bico},
  \citenamefont {Reyssat},\ and\ \citenamefont {Roman}}]{Bico2018}%
  \BibitemOpen
  \bibfield  {author} {\bibinfo {author} {\bibfnamefont {J.}~\bibnamefont
  {Bico}}, \bibinfo {author} {\bibfnamefont {{\'E}.}~\bibnamefont {Reyssat}}, \
  and\ \bibinfo {author} {\bibfnamefont {B.}~\bibnamefont {Roman}},\
  }\href@noop {} {\bibfield  {journal} {\bibinfo  {journal} {Annual Review of
  Fluid Mechanics}\ }\textbf {\bibinfo {volume} {50}},\ \bibinfo {pages} {629}
  (\bibinfo {year} {2018})}\BibitemShut {NoStop}%
\bibitem [{\citenamefont {Roman}\ and\ \citenamefont {Bico}(2010)}]{Roman2010}%
  \BibitemOpen
  \bibfield  {author} {\bibinfo {author} {\bibfnamefont {B.}~\bibnamefont
  {Roman}}\ and\ \bibinfo {author} {\bibfnamefont {J.}~\bibnamefont {Bico}},\
  }\href@noop {} {\bibfield  {journal} {\bibinfo  {journal} {Journal of
  Physics: Condensed Matter}\ }\textbf {\bibinfo {volume} {22}},\ \bibinfo
  {pages} {493101} (\bibinfo {year} {2010})}\BibitemShut {NoStop}%
\bibitem [{\citenamefont {Style}\ \emph {et~al.}(2020)\citenamefont {Style},
  \citenamefont {Tutika}, \citenamefont {Kim},\ and\ \citenamefont
  {Bartlett}}]{Style2020}%
  \BibitemOpen
  \bibfield  {author} {\bibinfo {author} {\bibfnamefont {R.~W.}\ \bibnamefont
  {Style}}, \bibinfo {author} {\bibfnamefont {R.}~\bibnamefont {Tutika}},
  \bibinfo {author} {\bibfnamefont {J.~Y.}\ \bibnamefont {Kim}}, \ and\
  \bibinfo {author} {\bibfnamefont {M.~D.}\ \bibnamefont {Bartlett}},\
  }\href@noop {} {\bibfield  {journal} {\bibinfo  {journal} {Advanced
  Functional Materials}\ ,\ \bibinfo {pages} {2005804}} (\bibinfo {year}
  {2020})}\BibitemShut {NoStop}%
\bibitem [{\citenamefont {Holmes}(2019)}]{Holmes2019}%
  \BibitemOpen
  \bibfield  {author} {\bibinfo {author} {\bibfnamefont {D.~P.}\ \bibnamefont
  {Holmes}},\ }\href@noop {} {\bibfield  {journal} {\bibinfo  {journal}
  {Current Opinion in Colloid \& Interface Science}\ }\textbf {\bibinfo
  {volume} {40}},\ \bibinfo {pages} {118 } (\bibinfo {year}
  {2019})}\BibitemShut {NoStop}%
\bibitem [{\citenamefont {Bico}\ \emph {et~al.}(2004)\citenamefont {Bico},
  \citenamefont {Roman}, \citenamefont {Moulin},\ and\ \citenamefont
  {Boudaoud}}]{Bico2004}%
  \BibitemOpen
  \bibfield  {author} {\bibinfo {author} {\bibfnamefont {J.}~\bibnamefont
  {Bico}}, \bibinfo {author} {\bibfnamefont {B.}~\bibnamefont {Roman}},
  \bibinfo {author} {\bibfnamefont {L.}~\bibnamefont {Moulin}}, \ and\ \bibinfo
  {author} {\bibfnamefont {A.}~\bibnamefont {Boudaoud}},\ }\href@noop {}
  {\bibfield  {journal} {\bibinfo  {journal} {Nature}\ }\textbf {\bibinfo
  {volume} {432}},\ \bibinfo {pages} {690} (\bibinfo {year}
  {2004})}\BibitemShut {NoStop}%
\bibitem [{\citenamefont {Elettro}\ \emph {et~al.}(2016)\citenamefont
  {Elettro}, \citenamefont {Neukirch}, \citenamefont {Vollrath},\ and\
  \citenamefont {Antkowiak}}]{Elettro2016}%
  \BibitemOpen
  \bibfield  {author} {\bibinfo {author} {\bibfnamefont {H.}~\bibnamefont
  {Elettro}}, \bibinfo {author} {\bibfnamefont {S.}~\bibnamefont {Neukirch}},
  \bibinfo {author} {\bibfnamefont {F.}~\bibnamefont {Vollrath}}, \ and\
  \bibinfo {author} {\bibfnamefont {A.}~\bibnamefont {Antkowiak}},\ }\href@noop
  {} {\bibfield  {journal} {\bibinfo  {journal} {Proceedings of the National
  Academy of Sciences}\ }\textbf {\bibinfo {volume} {113}},\ \bibinfo {pages}
  {6143} (\bibinfo {year} {2016})}\BibitemShut {NoStop}%
\bibitem [{\citenamefont {Grandgeorge}\ \emph {et~al.}(2018)\citenamefont
  {Grandgeorge}, \citenamefont {Krins}, \citenamefont {Hourlier-Fargette},
  \citenamefont {Laberty-Robert}, \citenamefont {Neukirch},\ and\ \citenamefont
  {Antkowiak}}]{Grandgeorge2018}%
  \BibitemOpen
  \bibfield  {author} {\bibinfo {author} {\bibfnamefont {P.}~\bibnamefont
  {Grandgeorge}}, \bibinfo {author} {\bibfnamefont {N.}~\bibnamefont {Krins}},
  \bibinfo {author} {\bibfnamefont {A.}~\bibnamefont {Hourlier-Fargette}},
  \bibinfo {author} {\bibfnamefont {C.}~\bibnamefont {Laberty-Robert}},
  \bibinfo {author} {\bibfnamefont {S.}~\bibnamefont {Neukirch}}, \ and\
  \bibinfo {author} {\bibfnamefont {A.}~\bibnamefont {Antkowiak}},\ }\href@noop
  {} {\bibfield  {journal} {\bibinfo  {journal} {Science}\ }\textbf {\bibinfo
  {volume} {360}},\ \bibinfo {pages} {296} (\bibinfo {year}
  {2018})}\BibitemShut {NoStop}%
\bibitem [{\citenamefont {Lau}\ \emph {et~al.}(2003)\citenamefont {Lau},
  \citenamefont {Bico}, \citenamefont {Teo}, \citenamefont {Chhowalla},
  \citenamefont {Amaratunga}, \citenamefont {Milne}, \citenamefont {McKinley},\
  and\ \citenamefont {Gleason}}]{Lau2003}%
  \BibitemOpen
  \bibfield  {author} {\bibinfo {author} {\bibfnamefont {K.~K.~S.}\
  \bibnamefont {Lau}}, \bibinfo {author} {\bibfnamefont {J.}~\bibnamefont
  {Bico}}, \bibinfo {author} {\bibfnamefont {K.~B.~K.}\ \bibnamefont {Teo}},
  \bibinfo {author} {\bibfnamefont {M.}~\bibnamefont {Chhowalla}}, \bibinfo
  {author} {\bibfnamefont {G.~A.~J.}\ \bibnamefont {Amaratunga}}, \bibinfo
  {author} {\bibfnamefont {W.~I.}\ \bibnamefont {Milne}}, \bibinfo {author}
  {\bibfnamefont {G.~H.}\ \bibnamefont {McKinley}}, \ and\ \bibinfo {author}
  {\bibfnamefont {K.~K.}\ \bibnamefont {Gleason}},\ }\href@noop {} {\bibfield
  {journal} {\bibinfo  {journal} {Nano Letters}\ }\textbf {\bibinfo {volume}
  {3}},\ \bibinfo {pages} {1701} (\bibinfo {year} {2003})}\BibitemShut
  {NoStop}%
\bibitem [{\citenamefont {Syms}\ \emph {et~al.}(2003)\citenamefont {Syms},
  \citenamefont {Yeatman}, \citenamefont {Bright},\ and\ \citenamefont
  {Whitesides}}]{Syms2003}%
  \BibitemOpen
  \bibfield  {author} {\bibinfo {author} {\bibfnamefont {R.~R.~A.}\
  \bibnamefont {Syms}}, \bibinfo {author} {\bibfnamefont {E.~M.}\ \bibnamefont
  {Yeatman}}, \bibinfo {author} {\bibfnamefont {V.~M.}\ \bibnamefont {Bright}},
  \ and\ \bibinfo {author} {\bibfnamefont {G.~M.}\ \bibnamefont {Whitesides}},\
  }\href@noop {} {\bibfield  {journal} {\bibinfo  {journal} {Journal of
  Microelectromechanical Systems}\ }\textbf {\bibinfo {volume} {12}},\ \bibinfo
  {pages} {387} (\bibinfo {year} {2003})}\BibitemShut {NoStop}%
\bibitem [{\citenamefont {De-Volder}\ and\ \citenamefont
  {Hart}(2013)}]{De-Volder2013}%
  \BibitemOpen
  \bibfield  {author} {\bibinfo {author} {\bibfnamefont {M.}~\bibnamefont
  {De-Volder}}\ and\ \bibinfo {author} {\bibfnamefont {A.~J.}\ \bibnamefont
  {Hart}},\ }\href@noop {} {\bibfield  {journal} {\bibinfo  {journal}
  {Angewandte Chemie International Edition}\ }\textbf {\bibinfo {volume}
  {52}},\ \bibinfo {pages} {2412} (\bibinfo {year} {2013})}\BibitemShut
  {NoStop}%
\bibitem [{\citenamefont {Hines}\ \emph {et~al.}(2016)\citenamefont {Hines},
  \citenamefont {Petersen}, \citenamefont {Lum},\ and\ \citenamefont
  {Sitti}}]{Hines2016}%
  \BibitemOpen
  \bibfield  {author} {\bibinfo {author} {\bibfnamefont {L.}~\bibnamefont
  {Hines}}, \bibinfo {author} {\bibfnamefont {K.}~\bibnamefont {Petersen}},
  \bibinfo {author} {\bibfnamefont {G.~Z.}\ \bibnamefont {Lum}}, \ and\
  \bibinfo {author} {\bibfnamefont {M.}~\bibnamefont {Sitti}},\ }\href@noop {}
  {\bibfield  {journal} {\bibinfo  {journal} {Advanced Materials}\ ,\ \bibinfo
  {pages} {1603483}} (\bibinfo {year} {2016})}\BibitemShut {NoStop}%
\bibitem [{\citenamefont {Kwok}\ \emph {et~al.}(2020)\citenamefont {Kwok},
  \citenamefont {Huang}, \citenamefont {Mastrangeli},\ and\ \citenamefont
  {Gracias}}]{Kwok2020}%
  \BibitemOpen
  \bibfield  {author} {\bibinfo {author} {\bibfnamefont {K.~S.}\ \bibnamefont
  {Kwok}}, \bibinfo {author} {\bibfnamefont {Q.}~\bibnamefont {Huang}},
  \bibinfo {author} {\bibfnamefont {M.}~\bibnamefont {Mastrangeli}}, \ and\
  \bibinfo {author} {\bibfnamefont {D.~H.}\ \bibnamefont {Gracias}},\
  }\href@noop {} {\bibfield  {journal} {\bibinfo  {journal} {Advanced Materials
  Interfaces}\ }\textbf {\bibinfo {volume} {7}},\ \bibinfo {pages} {1901677}
  (\bibinfo {year} {2020})}\BibitemShut {NoStop}%
\bibitem [{\citenamefont {Plateau}(1873)}]{Plateau1873}%
  \BibitemOpen
  \bibfield  {author} {\bibinfo {author} {\bibfnamefont {J.~A.~F.}\
  \bibnamefont {Plateau}},\ }\href@noop {} {\bibfield  {journal} {\bibinfo
  {journal} {Statique exp{\'e}rimentale et th{\'e}orique des liquides soumis
  aux seules forces mol{\'e}culaires, Gauthier-Villars}\ } (\bibinfo {year}
  {1873})}\BibitemShut {NoStop}%
\bibitem [{\citenamefont {Hutzler}\ \emph {et~al.}(2009)\citenamefont
  {Hutzler}, \citenamefont {Barry}, \citenamefont {Grasland-Mongrain},
  \citenamefont {Smyth},\ and\ \citenamefont {Weaire}}]{Hutzler2009}%
  \BibitemOpen
  \bibfield  {author} {\bibinfo {author} {\bibfnamefont {S.}~\bibnamefont
  {Hutzler}}, \bibinfo {author} {\bibfnamefont {J.}~\bibnamefont {Barry}},
  \bibinfo {author} {\bibfnamefont {P.}~\bibnamefont {Grasland-Mongrain}},
  \bibinfo {author} {\bibfnamefont {D.}~\bibnamefont {Smyth}}, \ and\ \bibinfo
  {author} {\bibfnamefont {D.}~\bibnamefont {Weaire}},\ }\href@noop {}
  {\bibfield  {journal} {\bibinfo  {journal} {Colloids and Surfaces A:
  Physicochemical and Engineering Aspects}\ }\textbf {\bibinfo {volume}
  {344}},\ \bibinfo {pages} {37} (\bibinfo {year} {2009})}\BibitemShut
  {NoStop}%
\bibitem [{\citenamefont {Mojdehi}\ \emph {et~al.}(2016)\citenamefont
  {Mojdehi}, \citenamefont {Tavakol}, \citenamefont {Royston}, \citenamefont
  {Dillard},\ and\ \citenamefont {Holmes}}]{Mojdehi2016}%
  \BibitemOpen
  \bibfield  {author} {\bibinfo {author} {\bibfnamefont {A.~R.}\ \bibnamefont
  {Mojdehi}}, \bibinfo {author} {\bibfnamefont {B.}~\bibnamefont {Tavakol}},
  \bibinfo {author} {\bibfnamefont {W.}~\bibnamefont {Royston}}, \bibinfo
  {author} {\bibfnamefont {D.~A.}\ \bibnamefont {Dillard}}, \ and\ \bibinfo
  {author} {\bibfnamefont {D.~P.}\ \bibnamefont {Holmes}},\ }\href@noop {}
  {\bibfield  {journal} {\bibinfo  {journal} {Extreme Mechanics Letters}\
  }\textbf {\bibinfo {volume} {9}},\ \bibinfo {pages} {237 } (\bibinfo {year}
  {2016})}\BibitemShut {NoStop}%
\bibitem [{\citenamefont {Seguin}\ and\ \citenamefont
  {Gondret}(2018)}]{Seguin2018}%
  \BibitemOpen
  \bibfield  {author} {\bibinfo {author} {\bibfnamefont {A.}~\bibnamefont
  {Seguin}}\ and\ \bibinfo {author} {\bibfnamefont {P.}~\bibnamefont
  {Gondret}},\ }\href@noop {} {\bibfield  {journal} {\bibinfo  {journal} {Phys.
  Rev. E}\ }\textbf {\bibinfo {volume} {98}},\ \bibinfo {pages} {012906}
  (\bibinfo {year} {2018})}\BibitemShut {NoStop}%
\bibitem [{\citenamefont {Kolb}\ \emph {et~al.}(2017)\citenamefont {Kolb},
  \citenamefont {Legu{\'e}},\ and\ \citenamefont
  {Bogeat-Triboulot}}]{kolb2017}%
  \BibitemOpen
  \bibfield  {author} {\bibinfo {author} {\bibfnamefont {E.}~\bibnamefont
  {Kolb}}, \bibinfo {author} {\bibfnamefont {V.}~\bibnamefont {Legu{\'e}}}, \
  and\ \bibinfo {author} {\bibfnamefont {M.-B.}\ \bibnamefont
  {Bogeat-Triboulot}},\ }\href@noop {} {\bibfield  {journal} {\bibinfo
  {journal} {Physical biology}\ }\textbf {\bibinfo {volume} {14}},\ \bibinfo
  {pages} {065004} (\bibinfo {year} {2017})}\BibitemShut {NoStop}%
\bibitem [{\citenamefont {Whyte}\ \emph {et~al.}(2017)\citenamefont {Whyte},
  \citenamefont {Haffner}, \citenamefont {Tanaka}, \citenamefont {Hjelt},\ and\
  \citenamefont {Hutzler}}]{Whyte2017}%
  \BibitemOpen
  \bibfield  {author} {\bibinfo {author} {\bibfnamefont {D.}~\bibnamefont
  {Whyte}}, \bibinfo {author} {\bibfnamefont {B.}~\bibnamefont {Haffner}},
  \bibinfo {author} {\bibfnamefont {A.}~\bibnamefont {Tanaka}}, \bibinfo
  {author} {\bibfnamefont {T.}~\bibnamefont {Hjelt}}, \ and\ \bibinfo {author}
  {\bibfnamefont {S.}~\bibnamefont {Hutzler}},\ }\href@noop {} {\bibfield
  {journal} {\bibinfo  {journal} {Colloids and Surfaces A: Physicochemical and
  Engineering Aspects}\ }\textbf {\bibinfo {volume} {534}},\ \bibinfo {pages}
  {112} (\bibinfo {year} {2017})}\BibitemShut {NoStop}%
\bibitem [{\citenamefont {Langlois}\ and\ \citenamefont
  {Hutzler}(2017)}]{Langlois2017}%
  \BibitemOpen
  \bibfield  {author} {\bibinfo {author} {\bibfnamefont {V.~J.}\ \bibnamefont
  {Langlois}}\ and\ \bibinfo {author} {\bibfnamefont {S.}~\bibnamefont
  {Hutzler}},\ }\href@noop {} {\bibfield  {journal} {\bibinfo  {journal}
  {Colloids and Surfaces A: Physicochemical and Engineering Aspects}\ }\textbf
  {\bibinfo {volume} {534}},\ \bibinfo {pages} {105} (\bibinfo {year}
  {2017})}\BibitemShut {NoStop}%
\bibitem [{\citenamefont {Reinelt}\ \emph {et~al.}(2001)\citenamefont
  {Reinelt}, \citenamefont {Boltenhagen},\ and\ \citenamefont
  {Rivier}}]{Reinelt2001}%
  \BibitemOpen
  \bibfield  {author} {\bibinfo {author} {\bibfnamefont {D.}~\bibnamefont
  {Reinelt}}, \bibinfo {author} {\bibfnamefont {P.}~\bibnamefont
  {Boltenhagen}}, \ and\ \bibinfo {author} {\bibfnamefont {N.}~\bibnamefont
  {Rivier}},\ }\href@noop {} {\bibfield  {journal} {\bibinfo  {journal} {The
  European Physical Journal E}\ }\textbf {\bibinfo {volume} {4}},\ \bibinfo
  {pages} {299} (\bibinfo {year} {2001})}\BibitemShut {NoStop}%
\bibitem [{\citenamefont {Frazier}\ \emph {et~al.}(2020)\citenamefont
  {Frazier}, \citenamefont {Jiang},\ and\ \citenamefont
  {Burton}}]{Frazier2020}%
  \BibitemOpen
  \bibfield  {author} {\bibinfo {author} {\bibfnamefont {S.}~\bibnamefont
  {Frazier}}, \bibinfo {author} {\bibfnamefont {X.}~\bibnamefont {Jiang}}, \
  and\ \bibinfo {author} {\bibfnamefont {J.~C.}\ \bibnamefont {Burton}},\
  }\href@noop {} {\bibfield  {journal} {\bibinfo  {journal} {Physical Review
  Fluids}\ }\textbf {\bibinfo {volume} {5}},\ \bibinfo {pages} {013304}
  (\bibinfo {year} {2020})}\BibitemShut {NoStop}%
\bibitem [{\citenamefont {Johnston}\ \emph {et~al.}(2014)\citenamefont
  {Johnston}, \citenamefont {McCluskey}, \citenamefont {Tan},\ and\
  \citenamefont {Tracey}}]{Johnston2014}%
  \BibitemOpen
  \bibfield  {author} {\bibinfo {author} {\bibfnamefont {I.}~\bibnamefont
  {Johnston}}, \bibinfo {author} {\bibfnamefont {D.}~\bibnamefont {McCluskey}},
  \bibinfo {author} {\bibfnamefont {C.}~\bibnamefont {Tan}}, \ and\ \bibinfo
  {author} {\bibfnamefont {M.}~\bibnamefont {Tracey}},\ }\href@noop {}
  {\bibfield  {journal} {\bibinfo  {journal} {Journal of Micromechanics and
  Microengineering}\ }\textbf {\bibinfo {volume} {24}} (\bibinfo {year}
  {2014})}\BibitemShut {NoStop}%
\bibitem [{\citenamefont {Elias}\ \emph {et~al.}(2007)\citenamefont {Elias},
  \citenamefont {Hutzler},\ and\ \citenamefont {Ferreira}}]{Elias2007}%
  \BibitemOpen
  \bibfield  {author} {\bibinfo {author} {\bibfnamefont {F.}~\bibnamefont
  {Elias}}, \bibinfo {author} {\bibfnamefont {S.}~\bibnamefont {Hutzler}}, \
  and\ \bibinfo {author} {\bibfnamefont {M.}~\bibnamefont {Ferreira}},\
  }\href@noop {} {\bibfield  {journal} {\bibinfo  {journal} {European Journal
  of Physics}\ }\textbf {\bibinfo {volume} {28}},\ \bibinfo {pages} {755}
  (\bibinfo {year} {2007})}\BibitemShut {NoStop}%
\bibitem [{\citenamefont {Miller}\ \emph {et~al.}(2014)\citenamefont {Miller},
  \citenamefont {Lazarus}, \citenamefont {Audoly},\ and\ \citenamefont
  {Reis}}]{miller2014}%
  \BibitemOpen
  \bibfield  {author} {\bibinfo {author} {\bibfnamefont {J.}~\bibnamefont
  {Miller}}, \bibinfo {author} {\bibfnamefont {A.}~\bibnamefont {Lazarus}},
  \bibinfo {author} {\bibfnamefont {B.}~\bibnamefont {Audoly}}, \ and\ \bibinfo
  {author} {\bibfnamefont {P.~M.}\ \bibnamefont {Reis}},\ }\href@noop {}
  {\bibfield  {journal} {\bibinfo  {journal} {Physical review letters}\
  }\textbf {\bibinfo {volume} {112}},\ \bibinfo {pages} {068103} (\bibinfo
  {year} {2014})}\BibitemShut {NoStop}%
\bibitem [{\citenamefont {Goldstein}(1980)}]{Goldstein}%
  \BibitemOpen
  \bibfield  {author} {\bibinfo {author} {\bibfnamefont {H.}~\bibnamefont
  {Goldstein}},\ }\href@noop {} {\emph {\bibinfo {title} {Classical
  Mechanics}}}\ (\bibinfo  {publisher} {Addison-Wesley},\ \bibinfo {year}
  {1980})\BibitemShut {NoStop}%
\bibitem [{\citenamefont {Lanczos}(1986)}]{Lanczos}%
  \BibitemOpen
  \bibfield  {author} {\bibinfo {author} {\bibfnamefont {C.}~\bibnamefont
  {Lanczos}},\ }\href@noop {} {\emph {\bibinfo {title} {The Variational
  Principles of Mechanics}}}\ (\bibinfo  {publisher} {Dover},\ \bibinfo {year}
  {1986})\BibitemShut {NoStop}%
\bibitem [{\citenamefont {Eddings}\ \emph {et~al.}(2008)\citenamefont
  {Eddings}, \citenamefont {Johnson},\ and\ \citenamefont
  {Gale}}]{Eddings2008}%
  \BibitemOpen
  \bibfield  {author} {\bibinfo {author} {\bibfnamefont {M.~A.}\ \bibnamefont
  {Eddings}}, \bibinfo {author} {\bibfnamefont {M.~A.}\ \bibnamefont
  {Johnson}}, \ and\ \bibinfo {author} {\bibfnamefont {B.}~\bibnamefont
  {Gale}},\ }\href@noop {} {\bibfield  {journal} {\bibinfo  {journal} {Journal
  of Micromechanics and Microengineering}\ }\textbf {\bibinfo {volume} {18}},\
  \bibinfo {pages} {067001} (\bibinfo {year} {2008})}\BibitemShut {NoStop}%
\bibitem [{\citenamefont {Landsberg}\ \emph {et~al.}(2009)\citenamefont
  {Landsberg}, \citenamefont {Farhadifar}, \citenamefont {Ranft}, \citenamefont
  {Umetsu}, \citenamefont {Widmann}, \citenamefont {Bittig}, \citenamefont
  {Said}, \citenamefont {Jülicher},\ and\ \citenamefont
  {Dahmann}}]{Landsberg2009}%
  \BibitemOpen
  \bibfield  {author} {\bibinfo {author} {\bibfnamefont {K.~P.}\ \bibnamefont
  {Landsberg}}, \bibinfo {author} {\bibfnamefont {R.}~\bibnamefont
  {Farhadifar}}, \bibinfo {author} {\bibfnamefont {J.}~\bibnamefont {Ranft}},
  \bibinfo {author} {\bibfnamefont {D.}~\bibnamefont {Umetsu}}, \bibinfo
  {author} {\bibfnamefont {T.~J.}\ \bibnamefont {Widmann}}, \bibinfo {author}
  {\bibfnamefont {T.}~\bibnamefont {Bittig}}, \bibinfo {author} {\bibfnamefont
  {A.}~\bibnamefont {Said}}, \bibinfo {author} {\bibfnamefont {F.}~\bibnamefont
  {Jülicher}}, \ and\ \bibinfo {author} {\bibfnamefont {C.}~\bibnamefont
  {Dahmann}},\ }\href@noop {} {\bibfield  {journal} {\bibinfo  {journal}
  {Current Biology}\ }\textbf {\bibinfo {volume} {19}},\ \bibinfo {pages}
  {1950} (\bibinfo {year} {2009})}\BibitemShut {NoStop}%
\bibitem [{\citenamefont {Burke}\ \emph {et~al.}(2021)\citenamefont {Burke},
  \citenamefont {Möbius}, \citenamefont {Hjelt}, \citenamefont {Ketoja},\ and\
  \citenamefont {Hutzler}}]{Burke2021}%
  \BibitemOpen
  \bibfield  {author} {\bibinfo {author} {\bibfnamefont {S.}~\bibnamefont
  {Burke}}, \bibinfo {author} {\bibfnamefont {M.}~\bibnamefont {Möbius}},
  \bibinfo {author} {\bibfnamefont {T.}~\bibnamefont {Hjelt}}, \bibinfo
  {author} {\bibfnamefont {J.}~\bibnamefont {Ketoja}}, \ and\ \bibinfo {author}
  {\bibfnamefont {S.}~\bibnamefont {Hutzler}},\ }\href@noop {} {\bibfield
  {journal} {\bibinfo  {journal} {SN Applied Sciences}\ }\textbf {\bibinfo
  {volume} {3}},\ \bibinfo {pages} {1} (\bibinfo {year} {2021})}\BibitemShut
  {NoStop}%
\bibitem [{\citenamefont {Hjelt}\ \emph {et~al.}(2021)\citenamefont {Hjelt},
  \citenamefont {Ketoja}, \citenamefont {Kiiskinen}, \citenamefont {Koponen},\
  and\ \citenamefont {Pääkkönen}}]{Hjelt2021}%
  \BibitemOpen
  \bibfield  {author} {\bibinfo {author} {\bibfnamefont {T.}~\bibnamefont
  {Hjelt}}, \bibinfo {author} {\bibfnamefont {J.~A.}\ \bibnamefont {Ketoja}},
  \bibinfo {author} {\bibfnamefont {H.}~\bibnamefont {Kiiskinen}}, \bibinfo
  {author} {\bibfnamefont {A.~I.}\ \bibnamefont {Koponen}}, \ and\ \bibinfo
  {author} {\bibfnamefont {E.}~\bibnamefont {Pääkkönen}},\ }\href@noop {}
  {\bibfield  {journal} {\bibinfo  {journal} {Journal of Dispersion Science and
  Technology}\ }\textbf {\bibinfo {volume} {0}},\ \bibinfo {pages} {1}
  (\bibinfo {year} {2021})}\BibitemShut {NoStop}%
\end{thebibliography}%
\end{document}